\documentclass[conf]{new-aiaa}
\usepackage[utf8]{inputenc}
\usepackage{graphicx}
\usepackage{xcolor}
\usepackage{amsmath}
\usepackage[version=4]{mhchem}
\usepackage{siunitx}
\usepackage{longtable,tabularx}
\usepackage{fixme}
\DeclareGraphicsExtensions{.ps}
\hypersetup{colorlinks=true}
\usepackage{tikz}
\usepackage{subcaption}

\newcommand{\Rey}{\mathcal{R}\mathfrak{e}}

\usepackage{hyperref}
\hypersetup{
    colorlinks=true,
    linkcolor=blue,
    filecolor=,      
    urlcolor=,
    pdftitle={Overleaf Example},
    pdfpagemode=FullScreen,
    }
\UseRawInputEncoding
\fxsetup{status=draft, layout=inline, theme=color}

\title{
Study of Dynamic Interaction Between Low Re Aerodynamic Load and Flexible-Biomimetic Wings with Tailorable Stiffness by FSI Modeling
}
\author{Smail Boughou\footnote{PhD Candidate, smail.boughou@uir.ac.ma , AIAA student member}}
\author{Ashraf A. Omar\footnote{Professor, ashraf.omar@uir.ac.ma, AIAA Senior Member}}
\author{Omer A. Elsayed\footnote{Professor, omer.almatbagi@uir.ac.ma}}
\affil{School of Aerospace and Automotive Engineering, LERMA Lab, Université Internationale de Rabat, Rocade Rabat Salé 11100, Rabat-Sala El Jadida, Morocco}
\author{Radouan Boukharfane
\footnote{Research \& Education Fellow, radouan.boukharfane@um6p.ma, AIAA Member}}
\affil{MSDA group, Mohammed VI Polytechnic University (UM6P), Benguerir, Morocco}
\author{Daniel J. Inman
\footnote{Harm Buning Collegiate Professor of Aerospace, daninman@umich.edu, AIAA Fellow}}
\affil{Department of Aerospace Engineering, University of Michigan, 1320 Beal Ave, Ann Arbor, Michigan, USA}
\begin{document}

\newcommand{\headernote}
\centering
\vspace*{-1.8cm}
\fboxrule=0.6pt \fboxsep=3pt
\fbox{\begin{minipage}{0.99\linewidth} 
This is the author's version of an article that has been published at 2023 AIAA SciTech Forum - 23-27 January 2023. The final version of record is available at: \url{https://arc.aiaa.org/doi/abs/10.2514/6.2023-0826}
\end{minipage}
}

\maketitle
\begin{abstract}
In the present work, we investigate dynamic interaction and response of flexible bio-inspired morphing wing structure to a low Reynolds aerodynamic load.
The aspects of inspiration are as follows. First, the segmentation of the wing into rigid and flexible segments.
Considering a leading edge constitution of bone and muscle.
In addition to a flexible trailing edge composed of feathers.
Second, the material properties provided by experimental biology in literature are adopted such as the bending stiffness and Young's modulus.
The development of numerical models allowing non uniform distribution of properties are developed and implemented into an OpenFoam finite volume solver that couples fluid dynamics to a structural solid dynamics solver through the FSI interface.
In the course of this work, the validation is performed for a NACA6409 airfoil considering a rigid segment of 40\% and flexible segment 60\% chord length in order to test the aero-structure behavior for an aerodynamic load of air flow at low Reynolds number $\boldsymbol{\Rey}$ of $\boldsymbol{5\times 10^5}$ for the fluid and feather inspired material properties.
The results suggest that bio-inspired techniques can be reproduced in engineering configurations.

\end{abstract}

\section{Nomenclature}
{\renewcommand\arraystretch{0.8}
\noindent\begin{longtable*}{@{}l @{\quad=\quad} l@{}}
$\mathrm{UAV}$        & Unmanned Aerial Vehicle \\
$\mathcal{U}_\infty$  & Freestream velocity \\
$c$                   & chord length \\
$\nu$                 & kinematic viscosity \\
$\Rey$                & Reynolds Number ($\Rey=\mathcal{U}_\infty c/\nu$) \\
$\mathcal{E}$         & Young's Modulus \\
$\Delta t$            & time step \\
$t^*$                 & time period \\
\end{longtable*}}

\section{Introduction}
\lettrine{T}he innovative ideas for bio-mimetic morphing micro aerial vehicles are promoted in the design of intelligent material structures and systems.
Interests are being gained to revisit the morphing concepts, due to the development of smart materials (such as piezo-electric composite).
Where the structure and the actuation device form a single mechanism.
Such a mechanism is needed for traditional airplanes. This unusual maneuver is seen in birds.

The comparison of the aerodynamic efficiency of gliding birds against UAVs is insightful for a better understanding of natural flight.
There has been an increased recognition that more attention needs to be paid to morphing wings. Wing morphing allows gulls to modulate static pitch stability during gliding \cite{Harvey2022GullMorphing}.
The morphing ability of the flexible membrane wing is provided by its flexibility, which allows it to adaptively alter the shape under aerodynamic loading.
The aerodynamic performance modeling and flow control are drawing the interest of zoologists, biologists, and the concerned aerodynamics community.
As a result, these researches combine the biological theory of natural flying with aerodynamic methodologies to address MAVs based on bird endurance.

Flexible wing is a successful way for improving the aerodynamic robustness of tiny fixed-wing drones operating in uncertain air situations by using a revolutionary biomimetic design.
The aim is to introduce a multidisciplinary approach to the study of biologically influenced flights by coupling aerodynamics, structure, and flight mechanics.

The bird wing aerodynamic is different in behavior compared to the conventional man-made wings as Withers \cite{Withers1981} found that the bird wings performed with low drag generally had low maximum lift coefficients, whereas wings with high maximum lift coefficients had high drag coefficients.

Their wings are models for the construction of noise-reducing applications \cite{Bachmann2010}.
Among the studied species is the owl. They are known for their silent flight because of the features in their wings that promote smooth flow \cite{jaworski2020,geyer2016}.
Thin and feather-like shapes that have a finite trailing edge thickness were designed by \citet{ananda2018aerodynamic} by using a multi-point inverse airfoil design technique in PROFOIL \cite{AirfoilDesignSoftwarefortheWeb} to design airfoil families (AS6091 to AS6099).
It consists of modifications in the finite trailing thickness between 4\%–6\% and can perform efficiently at the same bird Reynolds number scales ($10^4$-$10^5$).

\citet{Pabisch2010KeratinAlba} subjected the rachis to X-Ray, showing that both diameter and thickness decrease linearly from the base to the tip.

\begin{figure}[ht!]
\centering
\begin{subfigure}{.55\textwidth}
\centering
\includegraphics[width=2in, angle=90]{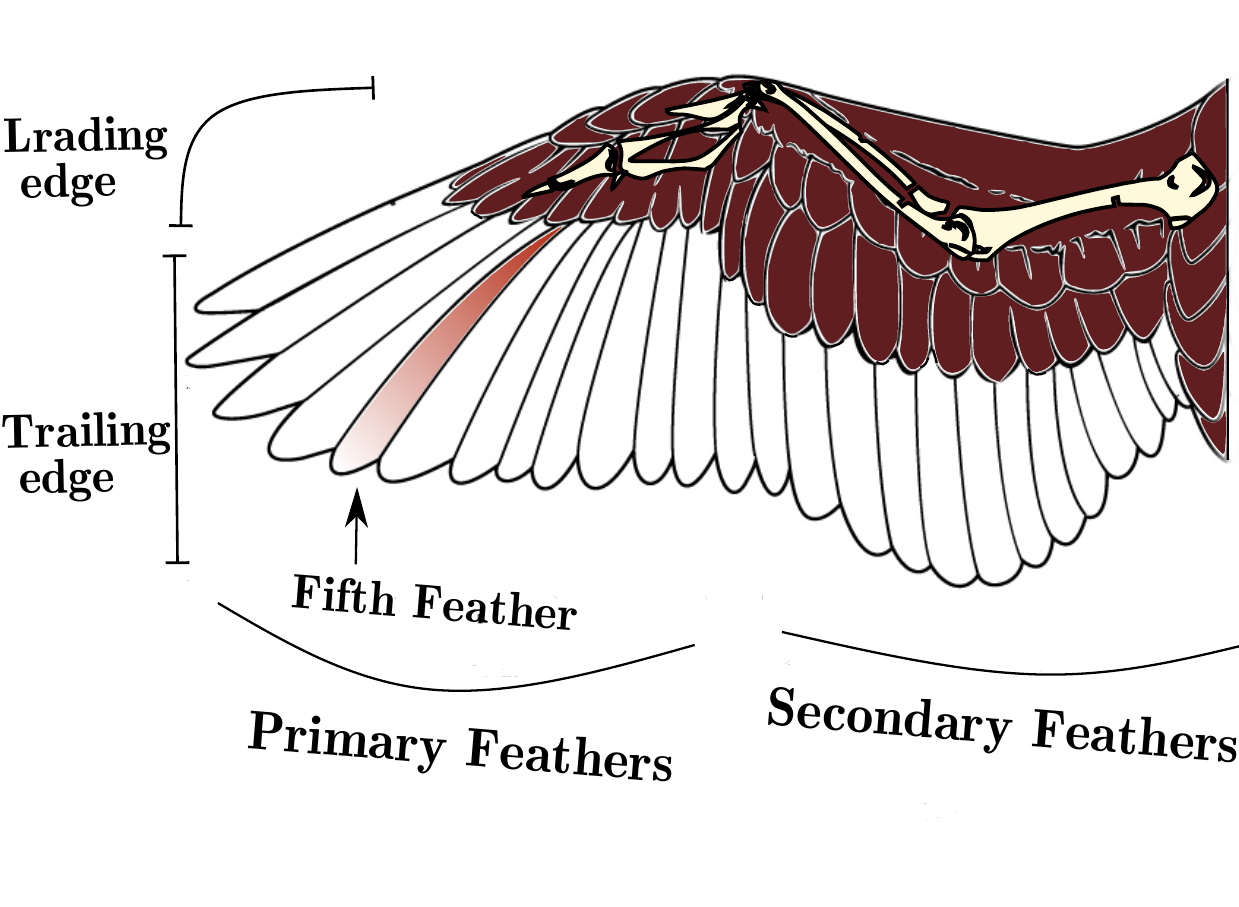}
\caption{wing anatomy}
\label{fig:anatomy} 
\end{subfigure}
\begin{subfigure}{.4\textwidth}
\centering
\includegraphics[width=.85\columnwidth]{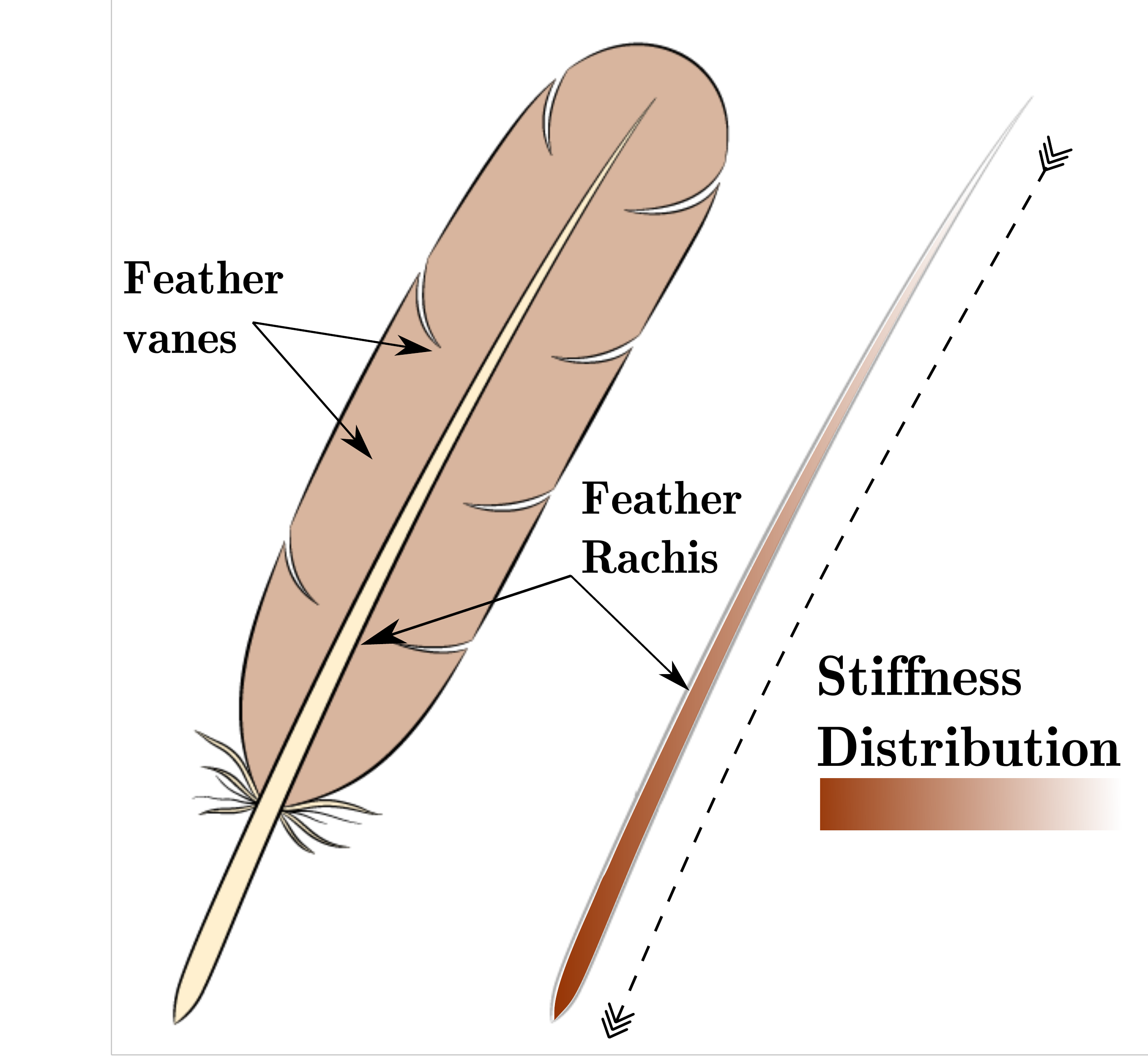}
\caption{Feather material properties inspiration}\label{fig:inspiration} 
\end{subfigure}
\caption{wing anatomy and Feather inspiration}
\label{fig:Ggeometry}
\end{figure}

\subsection{Morphing wings}\label{Morphing Wings}

Avian wing morphology allows inspiration for the aerostructural flow control and performance enhancement \cite{Wong2022FlexibleMorphing}.
It was found that dynamically synchronizing between structural deformations and flow vortices can improve aerodynamic performance.
The implementation of the biomimetic approach is meant to replicate the feather effects on aerodynamic performance.

Early studies on experimental biology focused on the material properties testing of the biological flights such as the wings and feathers structures \cite{Bachmann2012FlexuralProperties}.
Although some attempts have been made to address morphing wings, much of the work in this area is limited to steady CFD predictions of the original and morphing airfoils.
As a first step to understanding how the flow responds to dynamic morphing flap deflection, insightful work is presented by \citet{Abdessemed2018} using dynamic meshing to perform CFD analyses.
Here it is reported that those studies neglected the dynamic aspect of the interaction between fluid and wings.
In these directions, a growing field of researchers studying and developing avian-inspired morphing aircraft focused on the study of the morphing wings.
\citet{gamble2020load} found that the bio-inspired flexible airfoil maintained lift at Reynolds numbers below $1.5\times 10^5$, but at greater Reynolds numbers, the flexible airfoil alleviated the lift force and experienced trailing edge displacement.
It reduces the fluctuations of aerodynamic forces in a perturbed flow behind an oscillating plate by suppressing large-scale vortex shedding.
Previous work on understanding the low Re aerodynamic phenomena encountered while studying the owl-like airfoil \cite{Boughou2022} showed the unsteadiness of the aerodynamic coefficients.
The aero-structural response to the aerodynamic load is the motive behind the work presented in this paper.

The role of feather morphing hasn't been thoroughly investigated, and its impact on aerodynamics is unknown.
The aero-structural response of a flexible airfoil designed using biologically inspired structural and material data from feathers requires studies that concentrate on evaluating aerodynamic load and the aero-structural features in turbulent situations.
In the current study on bio-inspired flexible wings, we aim to make use of existing research in the field of experimental biology and complete several engineering objectives.
First, we examine bio-inspired structure requirements necessary to capture both flow field and structural analysis.
Unlike most of the state-of-the-art nonlinear aeroelastic frameworks that use the geometric nonlinear beam-based model, here finite volume solid mechanics is implemented.
Computational modeling can provide a way to develop predictive relationships between morphological traits and their impact on aerodynamic performance through a series of flexibility conditions that will be computed using Fluid-Structure interaction (FSI).
The primary objective of evaluations is to investigate the dynamic interaction between Low Re aerodynamic flow and Flexible-Biomimetic NACA6409 as earlier recommended in Ref \cite{gamble2020load} at an angle of attack $15^{\circ}$.
Following the work of \citet{gamble2020load}, we use simplified bird airfoil as the basis for this computational study and Young's modulus $\mathcal{E}$ of 2.5 GPa chordwise stiffness that may allow feather replication with a constant Young's modulus.
By doing so, this numerical testing of both the biological and the replicated flexible trailing edge verifies that these
bio-inspired techniques can be reproduced in engineering configurations.

\section{Material and Methods}

\subsection{Avian-like wing $\&$ Feather replication}

Bird wings cross section (airfoils) are of special profiles.
Thin and feather-like shapes having a finite trailing edge thickness are considered for the current work. 
Different biological flight aspects are inspired and mimicked. 
\subsubsection{mechanical properties inspiration}
Previous studies concluded that Young's modulus can be adjusted along the chord so that the equivalent stiffness distribution replicates the feather. Its shape was modelled as a cone \cite{Lingham-Soliar2017MicrostructuralFeathers}, whose thickness decreases linearly from the leading edge to the tip. This would provide a first insight into modeling material properties.

As \citet{Bachmann2012} reported in their experimental biology work, two parameters are mainly affecting flexibility of feathers.
They used two different methods (two-point bending test and Nanoindentation) to determine the Young's modulus of feather keratin of the barn owl which provided similar results.
They provided, in various biological experiments on the rachis, different material properties such as the Young's modulus $\mathcal{E}$ (Modulus of elasticity) and second moment of area $\mathcal{I}$ (second Moment of inertia).
Therefore, the flexural stiffness of the whole rachis is mainly sensitive to the cross-sectional geometry rather than by the Young's modulus of the keratin.
This latter ($\mathcal{I}$) dominates over the bending stiffness behavior along the feather length.
The variation in Young’s modulus along the shaft length is provided by experiment by \citet{Bonser1995TheKeratin}.
The feather shaft properties are examined by smaller samples that are analyzed to determine the variation of the tested properties along the feather length.
The measurement pattern shows to fit a linear regression (see Fig.\ref{fig:segments}).
Therefore, the variation of Young's modulus is supplied to the model as an expression of a linear function.
For stiffness, it decreases along the feather by five or more orders of magnitude.
Therefore, the tip almost vanishes in stiffness \cite{Bostandzhiyan2008FlexuralShaft}.
The same trend of increase in Young's modulus along the rachis was confirmed in \cite{Cameron2003YoungsFeathers}.

\begin{figure}[ht!]
\centering
\includegraphics[width=0.4\textwidth]{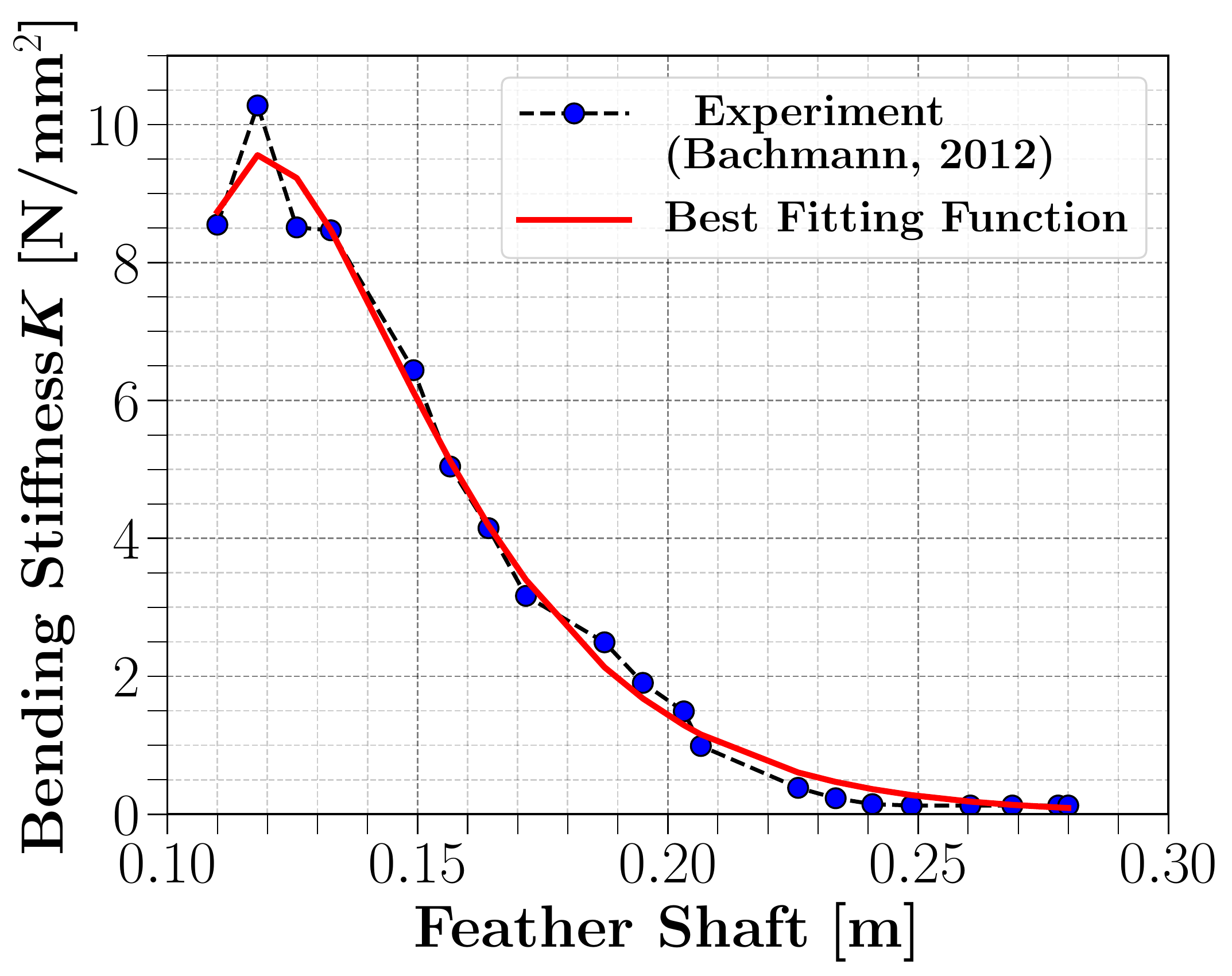}
\caption{Stiffness variation along the shaft length; Fitting with data from Ref \cite{Bachmann2012FlexuralProperties}}
\label{fig:stiffness}
\end{figure}

\subsection{Numerical Methods}
\subsubsection{rigid and flexible part inspiration}
As part of the validation of this work, this model builds upon the previous work of \citet{gamble2020load} in which the modeling of a feather-like airfoil is done by considering it as two segments: the rigid part and the flexible one.

From previous studies, bird-like low Reynolds number airfoils \cite{Boughou2022NumericalAirfoils} were shown to perform as a high lift airfoil.
Some conventional NACA airfoils exhibit such high lift in terms of aerodynamic performance.
The choice of NACA6409 for current investigations meets the bird wing cross-section analogy.

The rachis' structure in current study is mimicked with a wing cross section flexible at the trailing edge (cf. Fig.~\ref{fig:segments}).
Anatomy of bird airfoil considering rigid and flexible segments; NACA6409 airfoil segments (40\% ) rigid and (60\%) flexible denoted $\mathcal{L}$

\begin{figure}[ht!]
\centering
\includegraphics[width=0.5\textwidth]{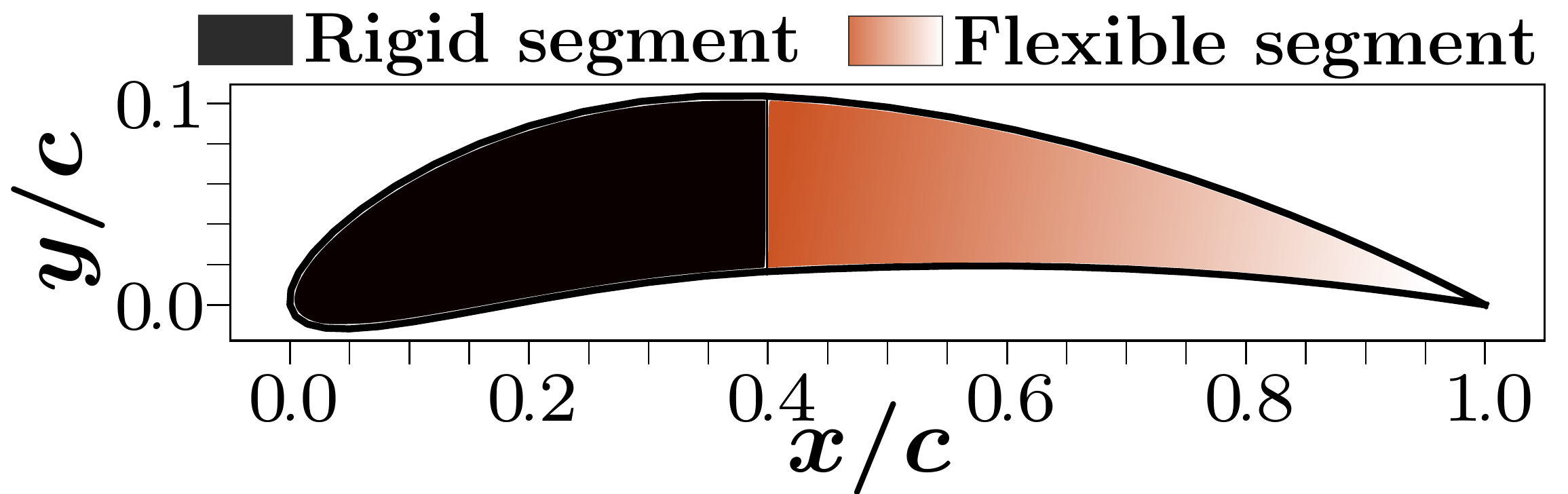}
\caption{Anatomy of bird airfoil considering rigid and flexible segments; NACA6409 airfoil segments (40\% ) rigid and (60\%) flexible denoted $\mathcal{L}$}
\label{fig:segments}
\end{figure}
\subsubsection{Wing Models}

\begin{figure}[ht!]
\centering
\begin{subfigure}{.5\textwidth}
\centering
\includegraphics[width=4in]{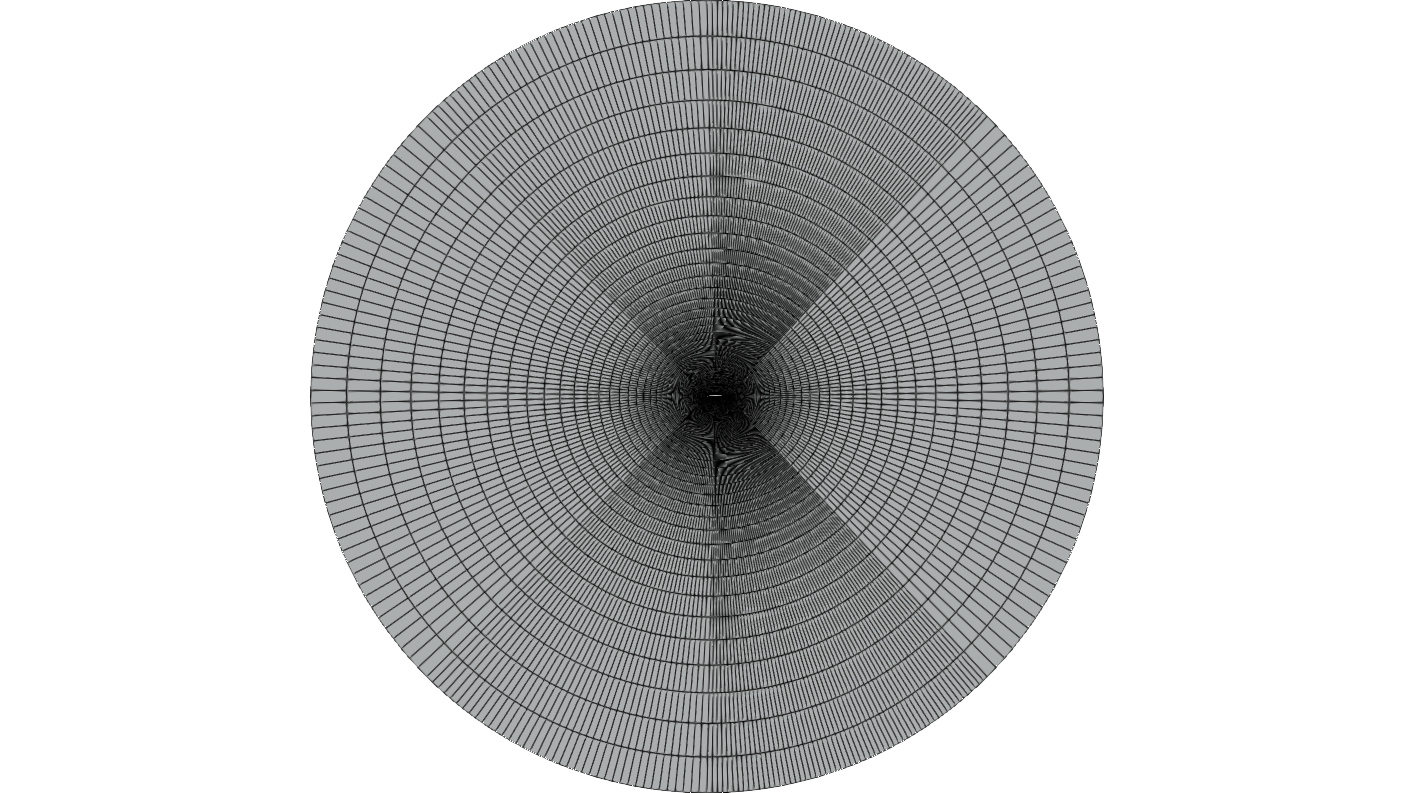}
\caption{\label{fig:Griddomain} Grid domain}
\label{fig:airfoildesigna}
\end{subfigure}
\begin{subfigure}{.4\textwidth}
\centering
\includegraphics[width=.6\columnwidth]{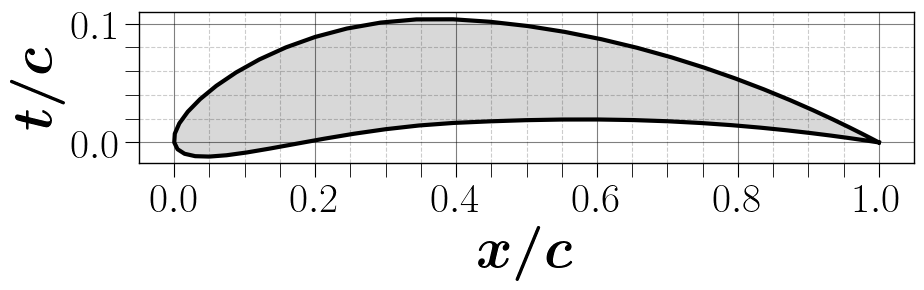}
\caption{\label{fig:owl}NACA6409}
\includegraphics[width=.6\columnwidth]{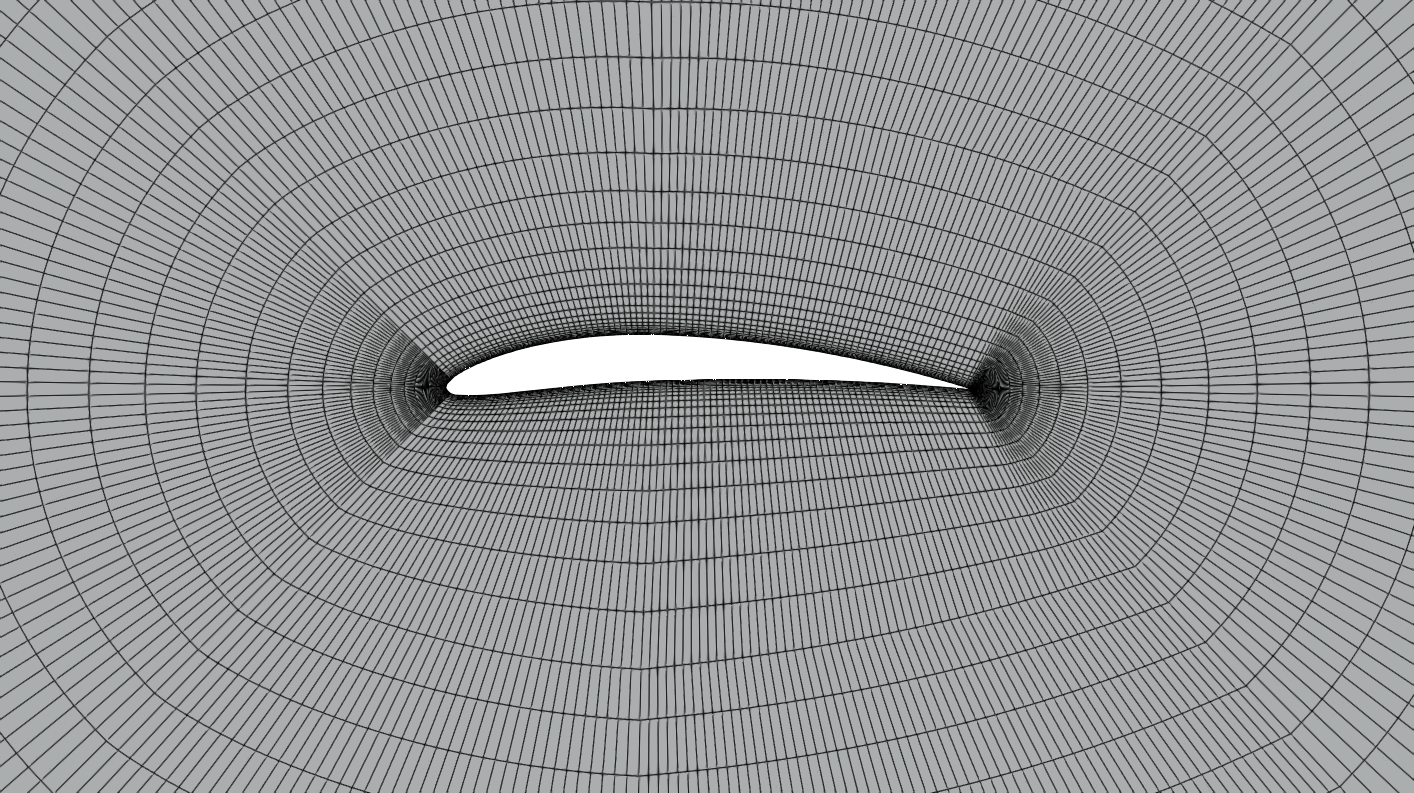}
\caption{\label{fig:6409}Grid around airfoil surface}
\end{subfigure}
\caption{Computational fluid domain and wing profile}
\label{fig:Ggeometry}
\end{figure}

The airfoil configuration and computational mesh used for the fluid–solid interaction problem is displayed in Fig.~\ref{fig:Ggeometry}.
As shown in Fig.~\ref{fig:Griddomain}, the computational domain is built with a structured grid using the ANSYS multi-blocks grid ICEM tool, with the appropriate boundary conditions.
The mesh plane is extruded in a span-wise direction with one cell.
The mesh is fine in order to avoid high aspect ratio cells within the domain.
The next section \ref{sec:results} outlines the preliminary resulting data assessed in multiple stages of the study of the NACA6409 airfoil (cf. Fig.~\ref{fig:6409}) at Reynolds number of $10^5$ to $5\times 10^5$ considering a rigid segment of 40\% and a flexible segment 60\% chord length (cf. Fig.~\ref{fig:airfoilSegments}).
The NACA6409 airfoil is defined to have a max camber of $ 6\%$ at $39.6\%$ chord at the location of the flexible trailing edge.

\begin{figure}[ht!]
\centering
\includegraphics[width=0.5\textwidth]{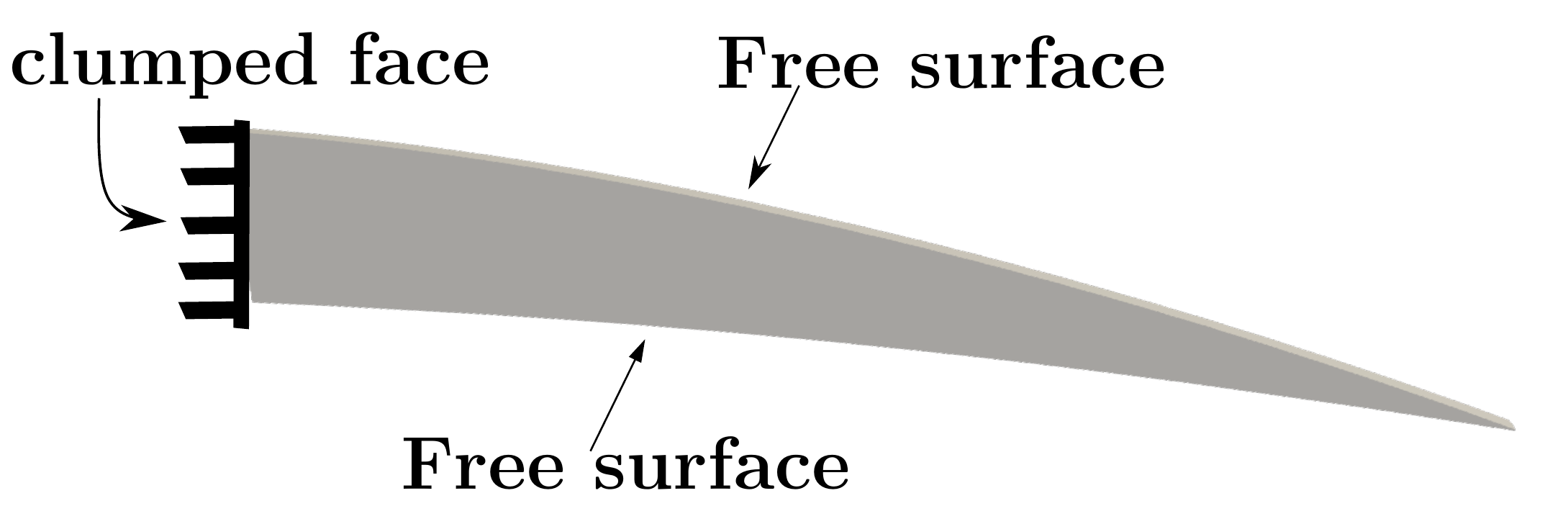}
\caption{ Boundary conditions for the flexible segment; NACA6409 airfoil segments (60\%) flexible }
\label{fig:segments}
\end{figure}

\subsubsection{Fluid-Structure Interaction}

Fluid-Structure Interaction (FSI) is defined as fluid flow applying forces to solid bodies and the solid’s dynamic response changes the surrounding flow field.
This is achieved through transferring information between the interfaces of the fluid and solid.
For the aero-elastic analysis of large-deformation structures, a nonlinear structural model has been developed, since the current biological materials do not exhibit linear elastic behavior.
To investigate how the structure responds to aerodynamic load, the governing equation for Neo-Hookean hyper-elasticity as described in \citet{Wiggins1998ComputationalInelasticity} is implemented and they read
\begin{equation}
\mu= \frac{\mathrm{E}}{2(1+\nu)}
\end{equation}
\begin{equation}
\mathcal{K}= \frac{\nu E}{(1+\nu)(1- 2\nu)} +\frac{2}{3}\mu
\end{equation}

FSI Simulations are performed with an open-source finite volume toolbox for solid mechanics and fluid-solid interaction simulations.
The finite volume method described for orthotropic bodies subjected to large strains and large deformations with consideration of updated Lagrangian finite volume solver \cite{Tukovic2014} is implemented in the FSI method described in this work.
To solve the well-known computational aerodynamics and aeroelastic airfoils, the FSI procedure is combined with the turbulence model k-$\omega$ SST and large deformation updated Lagrangian finite volume structure solver \cite{Cardiff2018}.

\begin{figure}[ht!]
\centering
\includegraphics[width=0.45\textwidth]{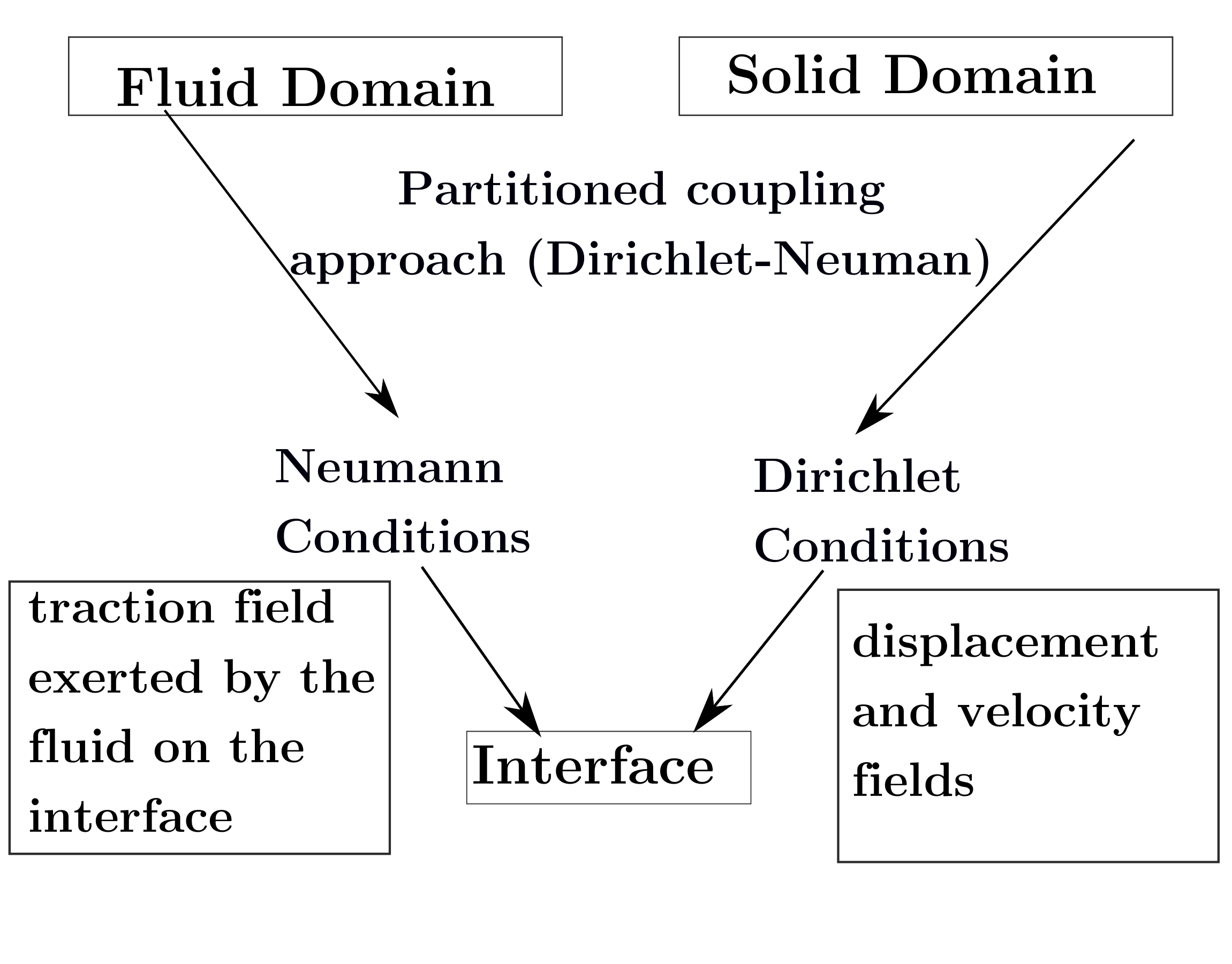}
\caption{FSI Coupling procedure}
\label{fig:airfoilSegments} 
\end{figure}

\section{FSI Analysis}
\label{sec:results}
\subsection{Flow Field}
\label{sec:flowField}
Comparison of the CFD results is done by quick assessment using XFoil panel method \cite{Drela1987}, it shows the stall of lift coefficient near an angle of attack 15$^\circ$.
The convergence of the steady state solution of the lift coefficient  at and AOA 15$^\circ$ is shown at Fig. \ref{fig:lfitCoe}
\begin{figure}[ht!]
\centering
\begin{subfigure}{.45\textwidth}
\includegraphics[width=0.99\columnwidth]{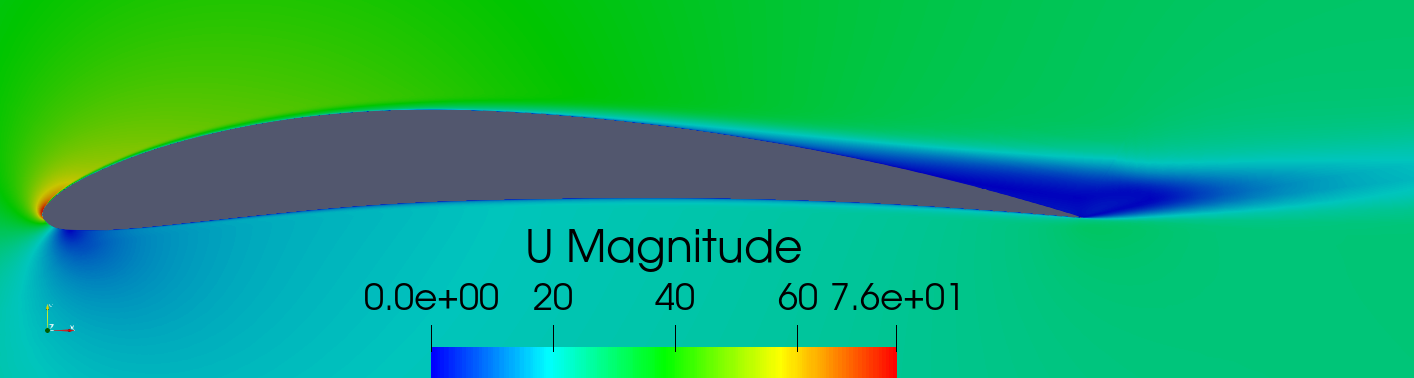}
\caption{Velocity contour of flow field at $\mathcal{R}_e=2\times10^5$}
\label{fig:velocitycontour}
\end{subfigure}
\begin{subfigure}{.45\textwidth}
\includegraphics[width=0.99\columnwidth]{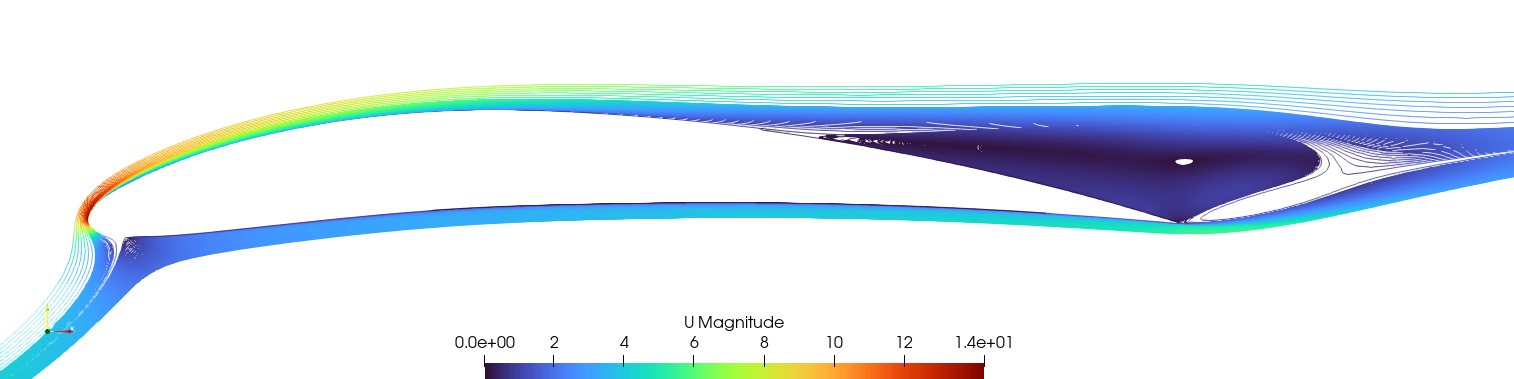}
\caption{Streamlines around airfoil obtained at $\mathcal{R}_e=2\times10^5$}
\label{fig:streamlines}
\end{subfigure}
\caption{\label{fig:history} Highlight of separation bubble at $\Rey =5\times10^5$ }
\label{fig:history}
\end{figure}

Within the stall region, the flow begins to separate over the upper surface. 
\begin{figure}[ht!]
\centering
\includegraphics[width=0.4\textwidth]{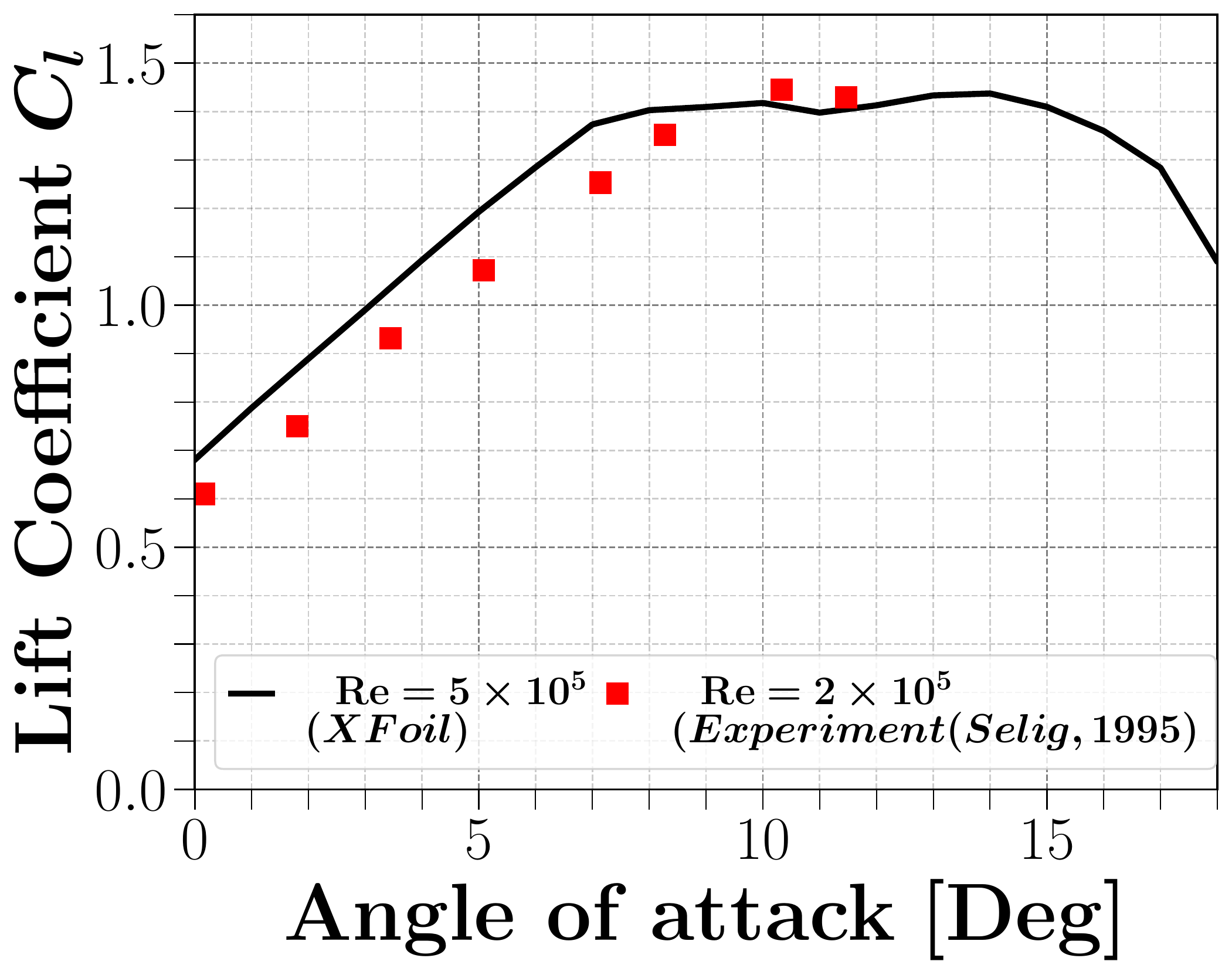}
\caption{Lift coefficient of NACA6409 airfoil at $\boldsymbol{\Rey$ = 5 $\times 10^5}$}
\label{fig:lfitCoe} 
\end{figure}

For reasons of ensuring flow development, the flow field around the NACA6094 at a $\mathcal{R_{\mathbf{e}}} = 5 \times 10 ^5$ is studied.
At the angle of attack $15^{\circ}$ is near-stall conditions where the flow begins to separate over the surface.
Which is the reason behind the choice of running the simulation at $15^{\circ}$.
The figure \ref{fig:lfitCoe} shows the non-dimensional time history of the aerodynamic lift coefficient ($\mathcal{C}_{\mathscr{l}}$) at the angle of attack $15^{\circ}$.
The resulting $\mathcal{C}_{\mathscr{l}}$ is in agreement with the experimental works of \citet{Selig1995} at University of Illinois at Urbana-Champaign (UIUC) wind tunnel tests.

\begin{figure}[ht!]
\centering
\includegraphics[width=0.4\textwidth]{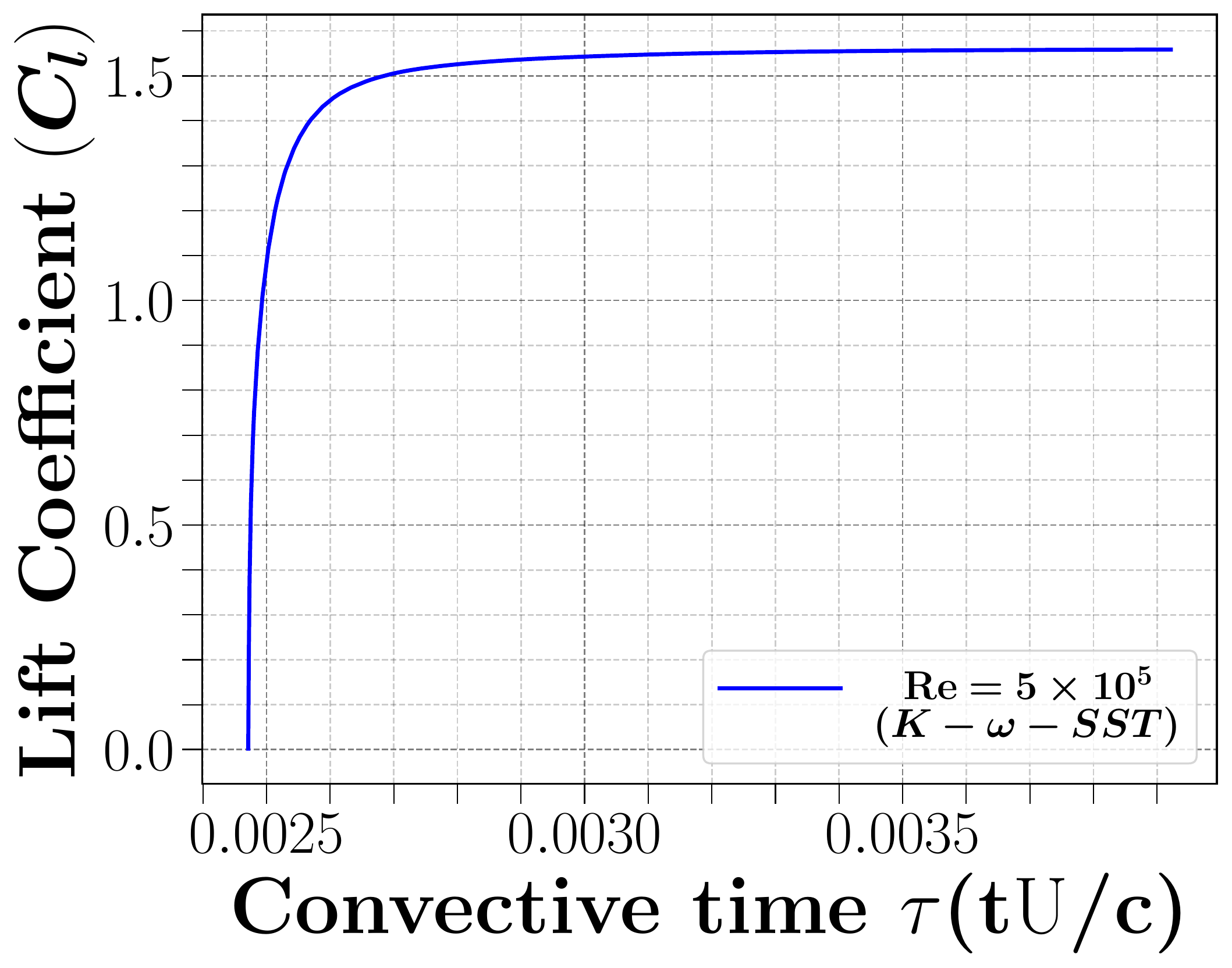}
\caption{Lift coefficient of NACA6409 airfoil at the angle of attack $15^{\circ}$ and $\boldsymbol{\Rey$ = 5 $\times 10^5}$}
\label{fig:lfitCoe} 
\end{figure}

\begin{figure}[ht!]
\centering
\includegraphics[width=0.4\textwidth]{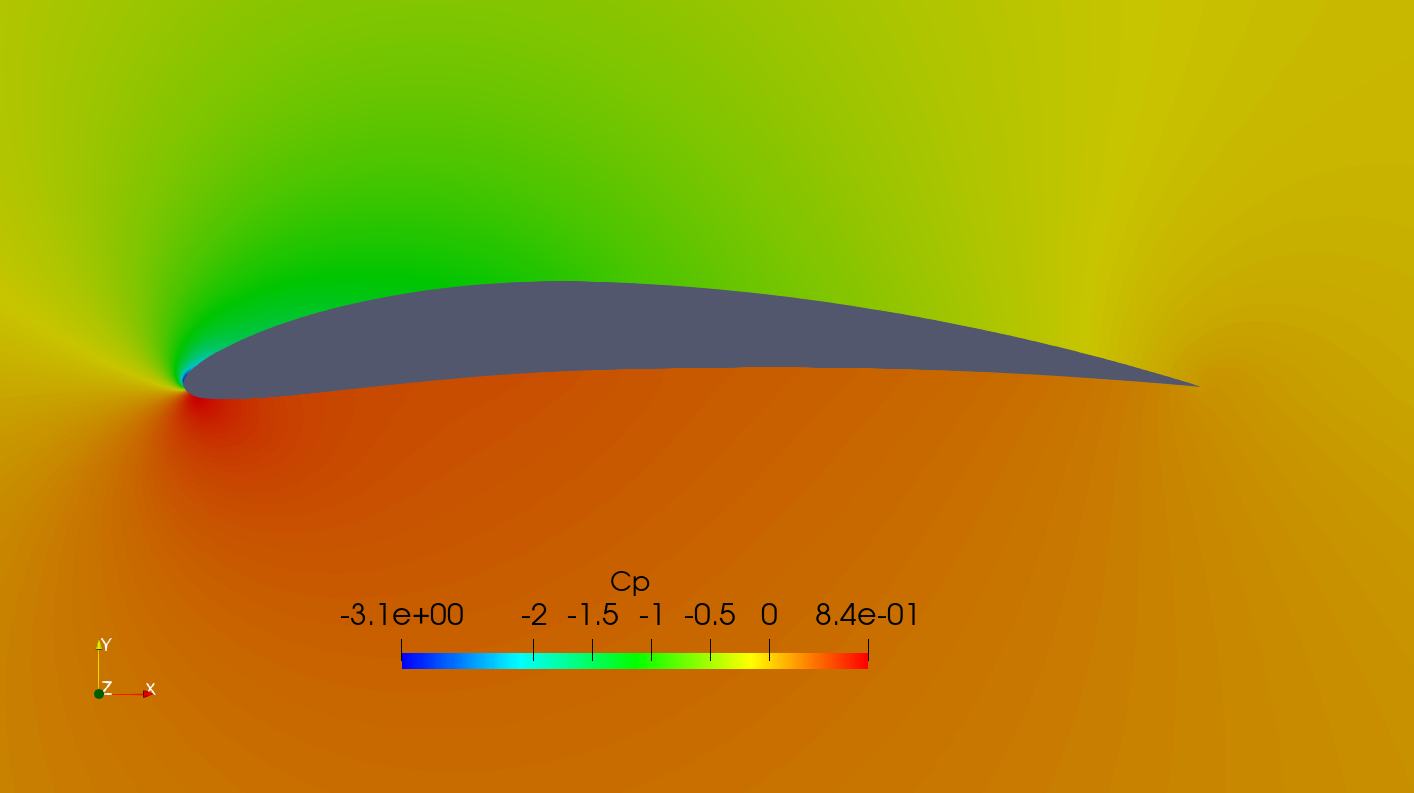}
\caption{Contour of pressure coefficient at and AOA 15 using $\boldsymbol{K-\omega-SST}$}
\label{fig:Cpcontour} 
\end{figure}

The difference of pressure over the upper and lower surfaces of the airfoil  shown in the figure \ref{fig:Cpcontour} will be applied in the solid surface.
\begin{figure}[ht!]
\centering
\includegraphics[width=0.4\textwidth]{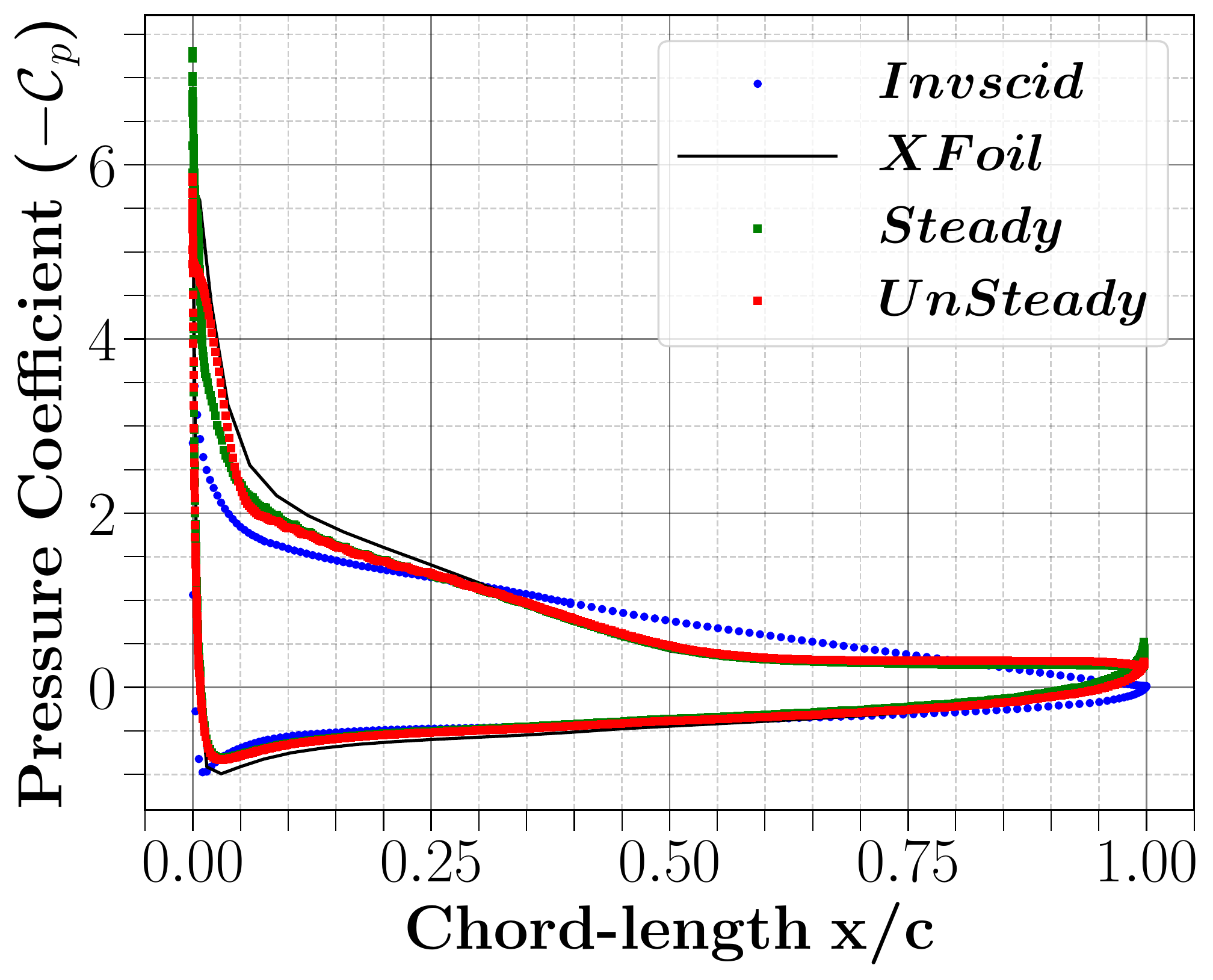}
\caption{Pressure Coefficient at and AOA $15^{\circ}$: XFoil vs CFD}
\label{fig:Cp} 
\end{figure}
%
\subsection{Structural Response}
\label{sec:structure}
As a preliminary assessment, a constant Young's modulus can indeed be used with reasonable thicknesses along the chord.
The purpose is to provide a strict verification benchmark and this test case serves more for demonstration purposes.
The tip and airfoil trailing edge deflection (cf. Sec. \S\ref{sec:deflection}) and fluid-structure response are described in the following sections (cf. Sec. \S\ref{sec:fluidStrucutreRespnseToYoungAndRey}).
\subsubsection{Tip deflection}
\label{sec:deflection}

\begin{figure}[ht!]
\centering

\includegraphics[width=0.5\columnwidth]{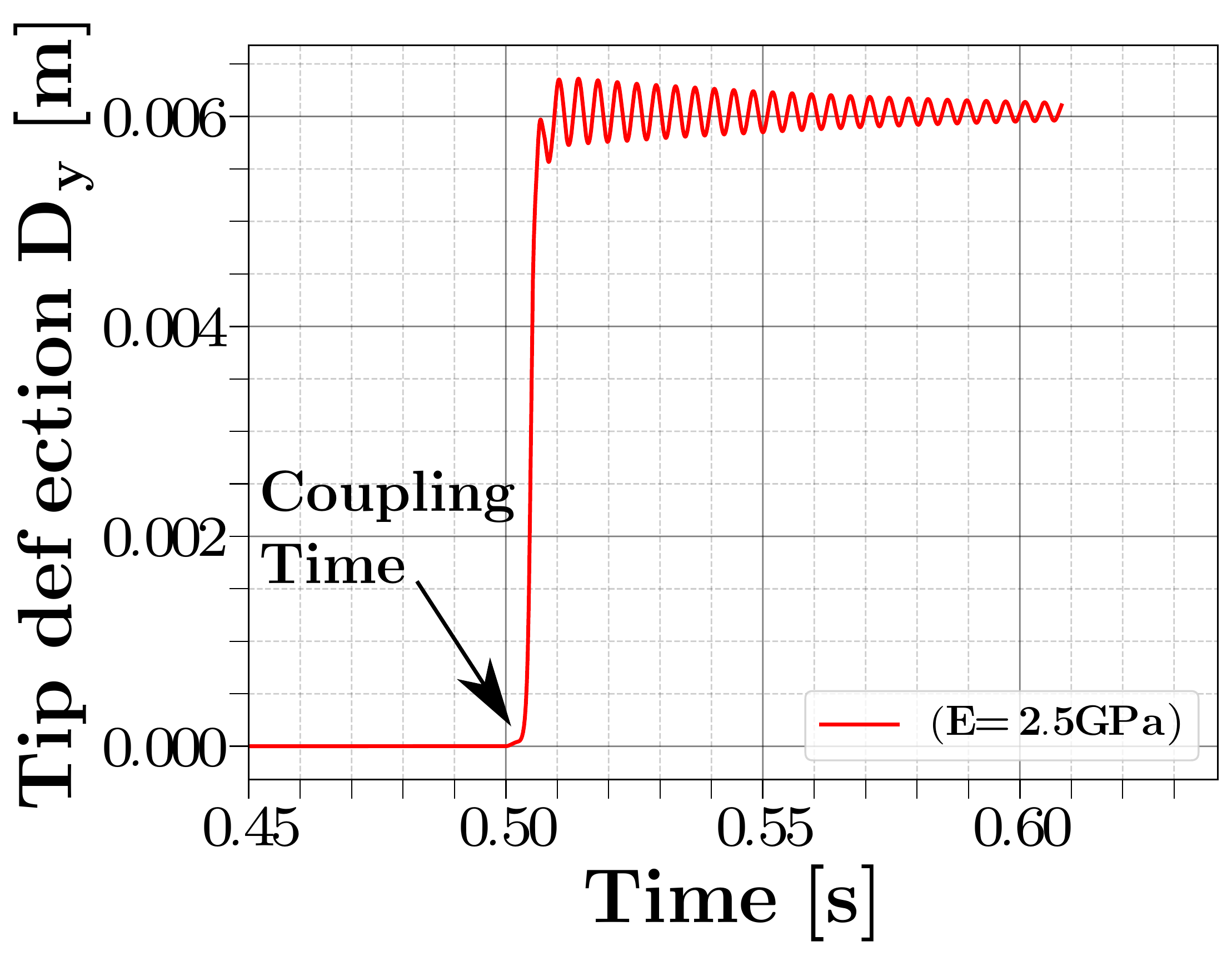}
\caption{ History of the tip displacement of the trailing edge at $\Rey =5\times10^5$ }
\label{fig:youngEffect}
\end{figure}

Here the results of the FSI implementation is illustrated for Young's modulus 2.5GPa and shown in Figure \ref{fig:youngEffect}.
The tip deflection history is shown in function of the displacement in the longitudinal direction ($\boldsymbol{D_y}$) for the tip control point on the trailing edge.
It illustrates the numerical computation of FSI which is done in two stages.
The standard approach for coupling fluid and structural solvers is to solve the fluid dynamic equations first, then transfer the computed loads to the structural equations.
First, the transient fluid flow is computed without coupling the fluid and structure interaction, considering the whole airfoil as a rigid body.
After the flow is developed along the wing, the system is coupled at a coupling time and takes into account the flexible segment of the airfoil.
In terms of the amount of deflection, a maximum value of deflection $10\%$ of the flexible part (we denote $\mathcal{L}$) (cf. Fig.~\ref{fig:youngEffect}).
However, the tip deflection is thought to remain in fluctuation motion.
Unlike the previous deformation, and for a Young's modulus $\mathcal{E}=2.5$ GPa, this displacement of the trailing edge tip is deflecting upwards in the same direction of the freestream velocity $\mathcal{U}_\infty$ with a maximum value of deflection $2.2\%$ of $\mathcal{L}$ (cf. Fig.~\ref{fig:airfoilSegments}).
The results from $\mathcal{E}=2.5$ GPa show a limited amount of fluctuations and demonstrate a state of slowdown.

The distribution of the displacement over the flexible segment $\mathcal{L}$ grows bigger towards the wing tip owing to large displacements at the tip for both values of $\mathcal{E}$.
Though, the lower $\mathcal{E}$ exhibits higher tip deflection.
Despite the differences, these displacement distributions provide support that the high $\mathcal{E}$ generates a sudden increase in deflection.

\begin{figure}[ht!]
\centering
\includegraphics[width=0.50\textwidth]{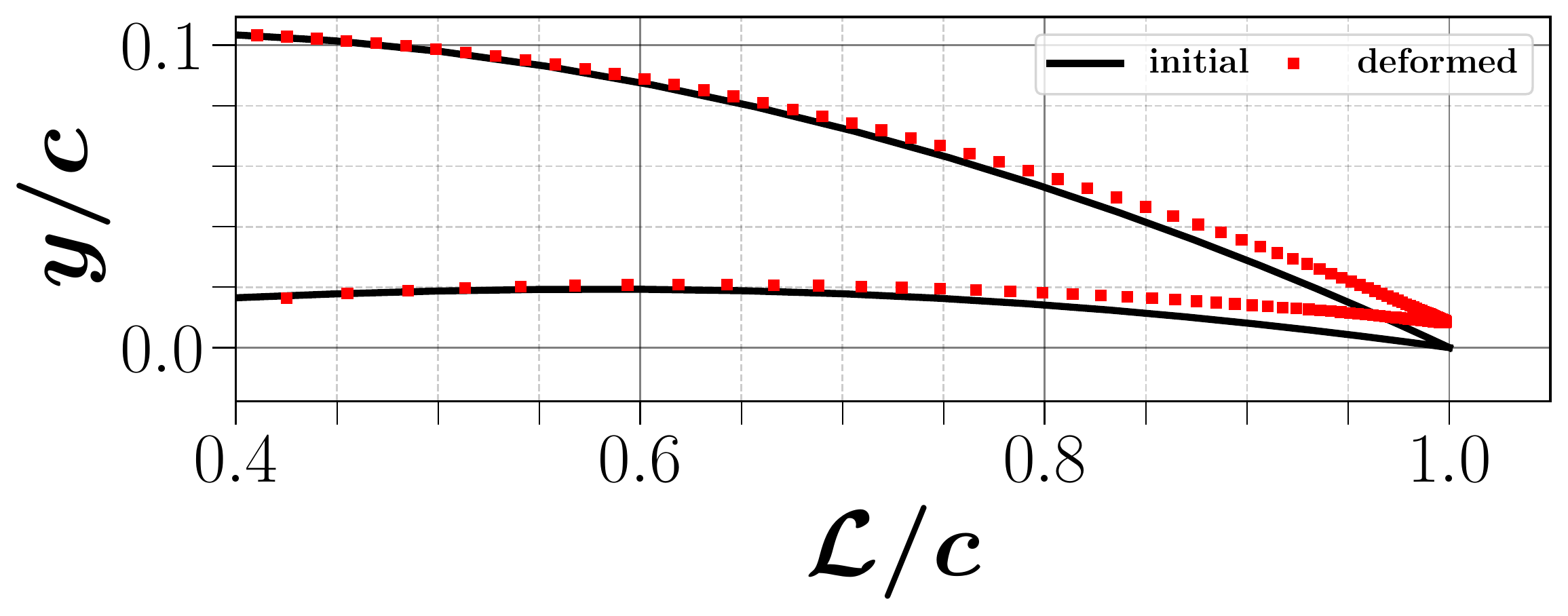}
\caption{NACA6409 airfoil flexible trailing edge deflection}
\label{fig:airfoilSegments} 
\end{figure}
\subsubsection{Fluid-structure Interaction}
\label{sec:fluidStrucutreRespnseToYoungAndRey}

\begin{figure}[ht!]
\centering
\includegraphics[width=0.45\columnwidth]{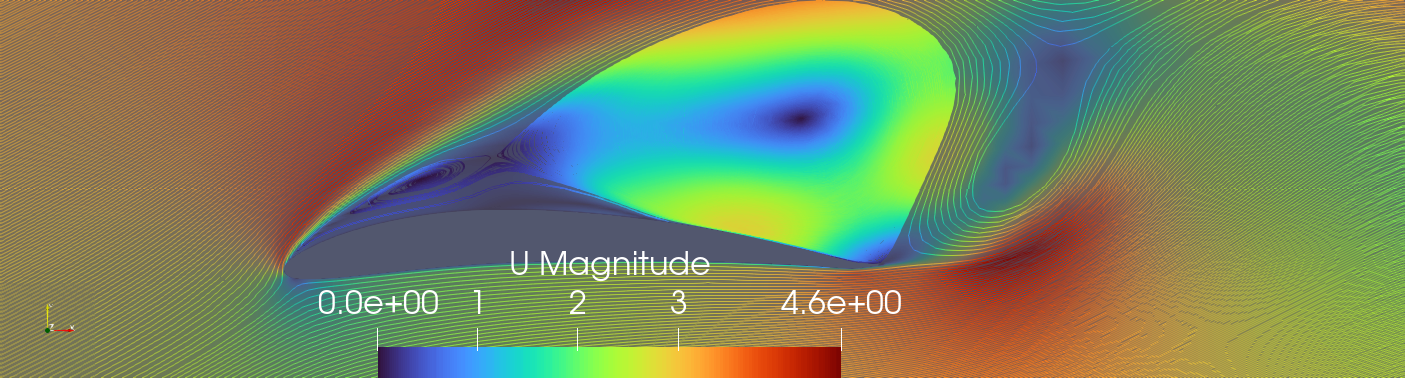}
\includegraphics[width=0.45\columnwidth]{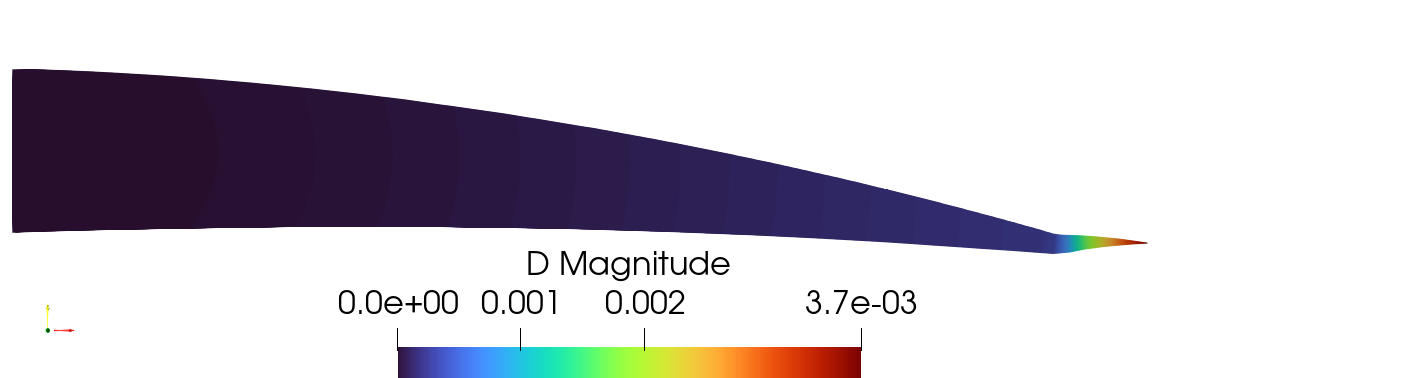}
\caption{Velocity colored flowfield streamlines (right) displacement contours (left) $\Rey=5\times10^5$ and $\mathcal{E}=2.5$ GPa}
\label{fig:25GPA}
\end{figure}

The computed instantaneous velocity colored streamlines are shown in Fig.~\ref{fig:25GPA} (left figure)  as well as the corresponding displacement magnitude contour of the flexible segment (right figure).
A small trailing edge bubble is formed and increases in diameter.
These separation bubbles come in a variety of sizes depending on the tip deflection.
For $\mathcal{E}=2.5$ GPa, it is not unexpected to see a small fluctuation of the tip and it tends to be stable with the fluid flow over the flexible segment. Therefore, the stable state of the tip deflection is highlighted and shows it relaxes towards a steady-state structure.

\begin{figure}[ht!]
\centering
\begin{subfigure}{.4\textwidth}
\includegraphics[width=0.99\columnwidth]{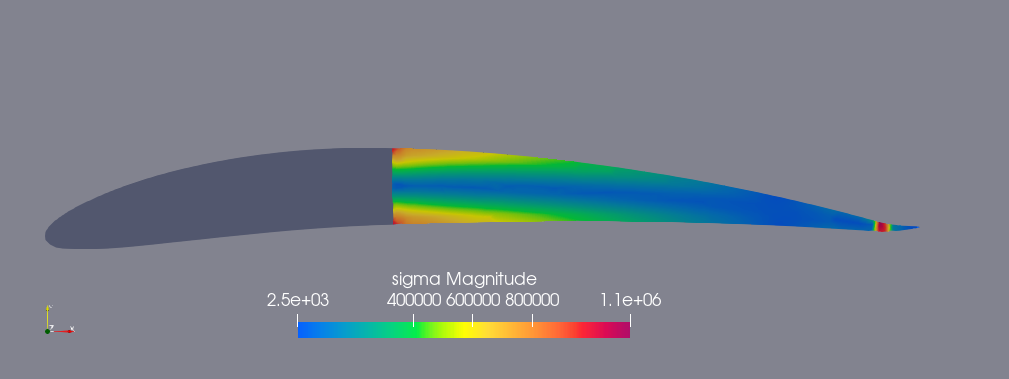}
\caption{\label{fig:stress} Equivalent stress contour over the flexible segment}
\end{subfigure}
\begin{subfigure}{.3\textwidth}
\includegraphics[width=0.99\columnwidth]{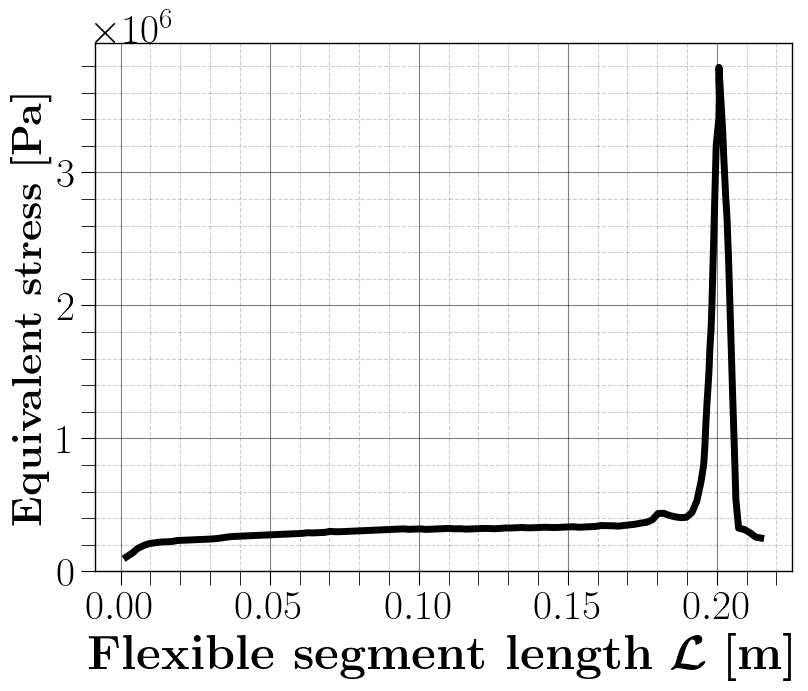}
\caption{Chordwise equivalent stress distribution\label{fig:stress2.5GPa}}
\end{subfigure}
\caption{\label{fig:displ} Stress analysis for $\mathcal{E}= 2.5$ GPa }
\end{figure}

Another goal of deformation control is to keep maximum wing deformation to a minimum in order to avoid damage from large stresses, which is critical for control in aero elastic applications, as it reduces material fatigue caused by vibrations and structural failures caused by high stresses.
The equivalent stress contour over the $\mathcal{L}$ segment and its distribution in chord wise direction are presented in Figs.~\ref{fig:stress} and \ref{fig:stress2.5GPa}, respectively.
The features learned from the distribution of equivalent stress over the $\mathcal{L}$ segment for $\mathcal{E}=2.5$ GPa shows promising stress relaxation with Young modulus.

\section{Conclusion}
Computational modeling can provide a way to develop predictive relationships between morphological traits and their impact on aerodynamic performance through a series of flexibility conditions that will be computed using Fluid-Structure interaction (FSI).
In this study, we have investigated the stable FSI for turbulent conditions for bio-inspired structure requirements necessary to capture both flow field and structural analysis.
One of the primary objectives of evaluations is to investigate the dynamic interaction between low Reynolds number aerodynamic flow and Flexible-Biomimetic NACA6409 as earlier recommended by \citet{gamble2020load}.
At an angle of attack of $15^{\circ}$, the results of the FSI implementation are illustrated for both Young's modulus $\mathcal{E}$ of 689.5 MPa and 2.5 GPa.
The study highlights the importance of understanding the stable mechanical properties sufficient for biomimetic adaptive structures and materials.
The results suggest that $\mathcal{E}=2.5$ GPa is a promising alternative to replicate the feather and will be used in further investigation of the bird-like wings.

\section*{Acknowledgments}
The Author S.B thanks Philip Cardiff collaboration for successfully implementing Openfoam Toolbox.
Thanks also to Jasmin Cheuk-Máhn Wong, Lawren Gamble and Christina Harvey for fruitful discussions on bird morphology.
This work is supported by the International University of Rabat (UIR).
The authors would like to thank the Simlab team for their support with the computing facilities of HPC Simlab-cluster of Mohammed VI Polytechnic University at Benguerir (UM6P).

\bibliography{main.bib}

\end{document}